# Attenuation of an Ultrasound Contrast Agent Estimated from Transient Solution of Linearized Rayleigh-Plesset Equation

Lang Xia

Email: lang4xia@gmail.com

**Abstract**

The attenuation of low-intensity acoustic waves in the suspension of ultrasound contrast agents (UCAs, microbubbles) is determined by the oscillation of the microbubbles in the medium. This bubble-induced attenuation is a linear phenomenon and can be estimated via a linearized Rayleigh-Plesset equation (RPE). In the material characterization, theoretical attenuation is estimated from steady state oscillation of an UCA and immediately compared with experimental attenuation data that are usually measured by shot-pulse ultrasound. However, discrepancy could exist in the characterization if the UCA does not ring up to steady state oscillation. In this article, we investigate the situation where the transient solution of the RPE is not negligible and discuss its impact on the modeling of the shell parameters of an UCA. We provide a formula for attenuation estimation considering the contribution due to transient oscillation.

**Keywords:** UCA, microbubbles, transient oscillation, acoustic attenuation

## I. INTRODUCTION

Microbubbles with encapsulated shells have been applied in the medical field as Ultrasound Contrast Agents (UCAs) for imaging purposes due to the compressibility of the gas core. The dynamic behaviors of microbubbles, governed by various types of Rayleigh-Plesset equations, dominate the contribution to the attenuation of acoustic waves in the suspension. A linear attenuation theory that is derived from linearized Rayleigh-Plesset equations is usually employed to determine the attenuation coefficient of the suspension of UCAs (Church 1995, Hoff, Sontum et al. 2000). Currently, two different approaches may be used to calculate the attenuation of UCA suspension. The first method, analogy of a driven harmonic oscillator, calculates the energy dissipation of a microbubble at linear oscillation and then summates the energy loss of each bubble in the unit volume to obtain the acoustic attenuation coefficient (Medwin 1977). The second method, assuming the suspension as an effective medium (a bubbly liquid), computes the attenuation of propagating acoustic waves in the medium using an effective wave equation along with a linearized Rayleigh-Plesset equation (Commander and Prosperetti 1989). These two methods produce identical results at low bubble volume fractions (Xia 2018).

The widely used attenuation formula is based on the steady state oscillation of a microbubble (Medwin 1977). As a forced linear harmonic oscillator, the microbubble takes a finite time to ring up to steady-state oscillation. Reduction in dissipation attributed to ''ring-up'' time effects was first observed by Akulichev (Akulichev, Bulanov et al. 1986). However, two more later studies were failed to reproduce the reduction results (Suiter 1992, Pace, Cowley et al. 1997). Clarke and Leighton addressed the failures and conducted a theoretical study into the time dependence of the response of air bubbles in fresh water to a continuous wave field, showing that the attenuation of a bubble is affected by cycle-length and amplitude of the driven pressure (Clarke and Leighton 2000). Akulichev and Bulanov also proposed a formula to estimate the scattering of bubbles in seawater due to transient oscillation (Akulichev and Bulanov 2011). It is well known that UCAs are coated microbubbles, of which the shell parameters are usually estimated through attenuation data that measured by broadband transducers (Sarkar, Shi et al. 2005, Gong, Cabodi et al. 2014, Xia, Paul et al. 2014). Depending on the pulse duration of the impinging ultrasound, the transient oscillation of an UCA could have significant impacts on the estimation of the attenuation coefficient. Since the accuracy of the attenuation formula is crucial in the characterization of shell materials, a reexamination of the attenuation due to transient oscillation is desired.

In this article, we investigate the attenuation of a single UCA driven by continuous sinusoidal pressure. By investigating the mechanical energy of the UCA, we obtained an equation that is capable of computing the attenuation coefficient of the UCA due to transient oscillation. This equation can also be used to model shell parameters. Comparisons between the results from linear attenuation theory and the present calculations are discussed.

## II. LINEAR ATTENUATION THEORY

The Rayleigh-Plesset equation describing the dynamics of a spherical microbubble can be written as

$$\rho\left(R\ddot{R}+\frac{3}{2}\dot{R}^2\right)+\frac{2\gamma}{R}+4\mu\frac{\dot{R}}{R}=p_g-p_\infty \tag{1}$$

where $\rho$ is the density of the surrounding liquid, $R$ is the instantaneous radius of the bubble and $\dot{R} = \partial R/\partial t$, $\ddot{R} = \partial^2 R/\partial t^2$, $\gamma$ is the surface tension, $\mu$ is the viscosity of the host liquid, $p_g$ is the pressure inside the bubble, and $p_\infty$ is the ambient pressure in the liquid. An UCA is a microbubble encapsulated by a shell material. Various materials, such as lipids, polymers, and proteins, can be used to encapsulate a free bubble, giving rise to different types of Rayleigh-Plesset equations. Since the shell material is not the focus for the present study, we simply assume the shell of an UCA to be viscoelastic (Sarkar, Shi et al. 2005). The dynamical equation for the UCA of equilibrium radius $R_0$, undergoing forced linear spherical pulsations ($R(t) = R_0 + X(t)$ and $|X(t)| \ll R_0$) at an external excitation pressure can be written in the form of

$$\ddot{X} + \omega_0 \delta \dot{X} + \omega_0^2 X = F\cos(\omega t) \tag{2}$$

with

$$\omega_0^2 = \frac{1}{\rho R_0^2}(3\kappa P_0 - \frac{4\gamma_0}{R_0} + \frac{4E^s}{R_0}) \tag{3}$$

$$\delta_l = \frac{4\mu}{\omega_0 R_0^2 \rho}, \quad \delta_s = \frac{4\kappa^s}{\omega_0 R_0^3 \rho}, \quad F = \frac{P_A}{\rho R_0} \tag{4}$$

where $X$ is the displacement around equilibrium radius $R_0$, $\omega_0$ is the bubble's pulsation angular resonance frequency, $\kappa$ is the polytropic constant, $P_0$ is the amplitude of the ambient pressure, $\gamma_0$ is a reference value of the interfacial tension. $\delta = \delta_l + \delta_s$ is the damping constant, of which the terms in the right-hand side stand for dampings of liquid viscosity and interface, respectively, $\kappa^s$ and $E^s$ are the dilatational viscosity and elasticity of the bubble shell, respectively, $P_A$ is the amplitude of the excitation pressure. Eq.(2) indicates that the linear dynamics of a coated microbubble (UCA) is a linear harmonic oscillator, having a full solution of the form.

$$X(t) = X_{st}(t) + X_{tr}(t) \tag{5}$$

where the steady-state solution is in the form of

$$X_{st} = \frac{F}{\left[(\omega_0^2 - \omega^2)^2 + (\delta\omega_0\omega)^2\right]^{1/2}} \cos(\omega t - \phi)$$

$$\phi = \tan^{-1}\left(\frac{\delta\omega_0\omega}{\omega_0^2 - \omega^2}\right)$$

(6)

and the transient solution is

$$X_{tr} = Ae^{-\delta\omega t/2}\cos(\omega_d t) + Be^{-\delta\omega t/2}\sin(\omega_d t) \tag{7}$$

with

$$A = -\frac{F}{\left[(\omega_0^2 - \omega^2)^2 + (\delta\omega_0\omega)^2\right]^{1/2}}\cos(\phi)$$

$$B = -\frac{F}{\left[(\omega_0^2 - \omega^2)^2 + (\delta\omega_0\omega)^2\right]^{1/2}} \frac{1}{\omega_d}\left[\omega\sin(\phi) + \frac{\delta\omega_0}{2}\cos(\phi)\right] \tag{8}$$

$$\omega_d = \omega_0\left(1 - \frac{\delta^2}{4}\right)^{1/2}$$

The damped angular resonance frequency $\omega_d$ may be complex values depending on the magnitude of the damping constant $\delta$. For the case of an UCA (coated microbubble), the damping constant is smaller than 1, and $\omega_d$ is a real-valued constant. This indicates that the UCA is underdamped, and the transient solution is decaying oscillatory with a frequency $\omega_d$. Before the transient oscillation decays to zero, it interferes with the steady state oscillation, generating perplexed bubble motions. The full and steady state solutions at a driven pressure of 40 kPa and excitation frequencies of *0.5f₀, f₀*, and *2f₀* are illustrated in **Figure 1**. The shell parameters for simulating the curves are radius $R_0 = 2.6\times10^{-6}$ m, the dilatational viscosity $\kappa^s = 1.97\times10^{-8}$ N.s/m, the dilatational elasticity $E^s = 0.4$ N/m, and the reference surface tension $\gamma_0 = 0.04$ N/m. The above values are typical shell parameters of a lipid encapsulated UCA at linear oscillations (Xia, Porter et al. 2015). These figures show that the amplitude differences of the oscillation curves are significant in the transient region before 1 µs. After that, the full and steady state oscillations are almost the same. Thus, one can expect that the attenuation calculated based on the steady state and transient oscillations will be different.

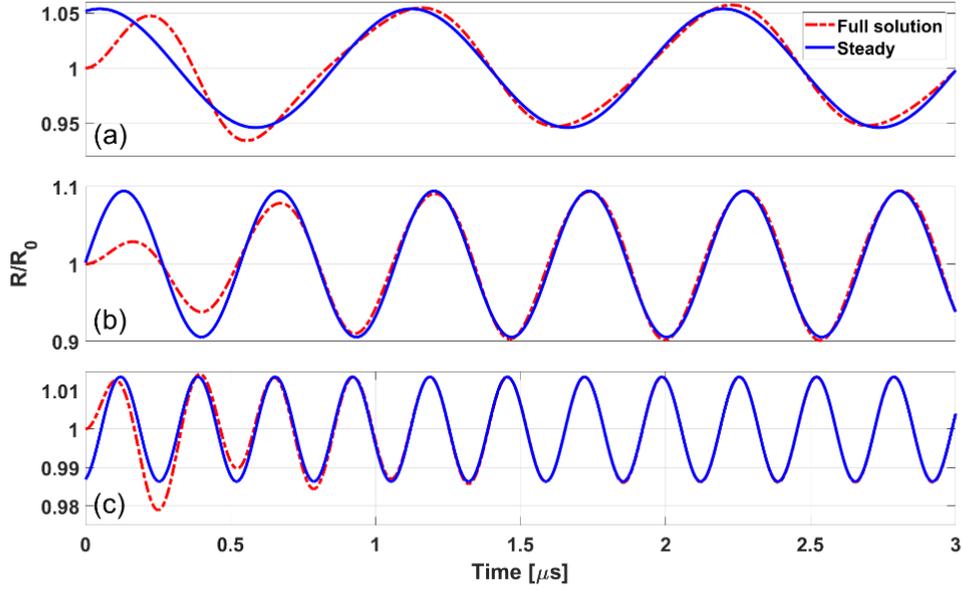

**Figure 1**: The oscillation of an UCA vs. excitation frequencies at (a) 0.5$f_0$, (b) $f_0$, and (c) 2$f_0$.

To see the influence of transient oscillation on the attenuation, we can investigate the energy absorption of a bubble in the acoustic field (Xia 2018). By assuming that an UCA oscillates spherically without thermal dissipation in an incompressible liquid, the energy delivered to the UCA can be written as

$$\Pi_i = (4\pi R^2 p)\dot{R} \qquad (9)$$

where $p = P_A \cos(\omega t)$ is the driven pressure. The average power delivered into the system is

$$\Pi = \frac{1}{T}\int_0^T \Pi_i dt \qquad (10)$$

where $T = 2\pi/\omega$ is the period. The average intensity of the impinging acoustic wave can be written as

$$I = \frac{P_A^2}{2\rho c_0} \qquad (11)$$

where $c_0$ is the sound speed in the host liquid, and the extinction cross-section is given by

$$\sigma_e = \frac{\Pi}{I} = \frac{4\omega\rho c_0}{P_A} \int_0^{2\pi/\omega} R^2 \dot{R} \cos(\omega t) dt \qquad (12)$$

The above equation is the extinction cross-section of a microbubble. Here we only investigate the effects of bubble oscillation on the energy attenuation, detail discussions on other mechanisms of dissipation may refer to a comprehensive review (Ainslie and Leighton 2011). By substituting the steady state solution Eq.(6) into Eq.(12), a linear attenuation theory in terms of the extinction cross-section $\sigma_e^{st}$ is given by the following equation (Medwin 1977, Xia 2018)

$$\sigma_e^{st} = 4\pi R_0^2 \frac{\delta c_0}{\omega_0 R_0} \frac{\Omega^2}{\left(1-\Omega^2\right)^2 + \delta^2 \Omega^2} \qquad (13)$$

where $\Omega = \omega/\omega_0$. The attenuation coefficient $\alpha$ of a bubble suspension can be computed readily by assuming a linear attenuation law, of which the final result is in the form of

$$\alpha = (20\log_{10} e)\frac{1}{2} n \sigma_e^{st} \text{ [dB/m}^2\text{]} \qquad (14)$$

where $e$ is the base of the natural logarithm, and $n$ is the total number of microbubbles per unit volume. Since the attenuation coefficient is proportional to the extinction cross-section, $\sigma_e$ or $\sigma_e^{st}$ is used to represent the attenuation as a matter of convenience.

By substituting Eq.(5) into Eq.(12), a general attenuation formula can be derived in terms of the full solution. This formula contains the contribution due to both transient and steady state solutions. Outside the transient region, **Figure 2a** shows that the attenuation calculated from the full solution is almost identical to the steady state attenuation (Medwin's formula Eq.(13)). Little difference of the attenuation peaks exists at the resonance frequency. However, inside the transient region, **Figure 2b** indicates that the peak attenuation calculated from the full solution is about 2/3 of the steady state attenuation, and its location shifts to the damped resonance frequency, slightly before the resonance frequency of the UCA.

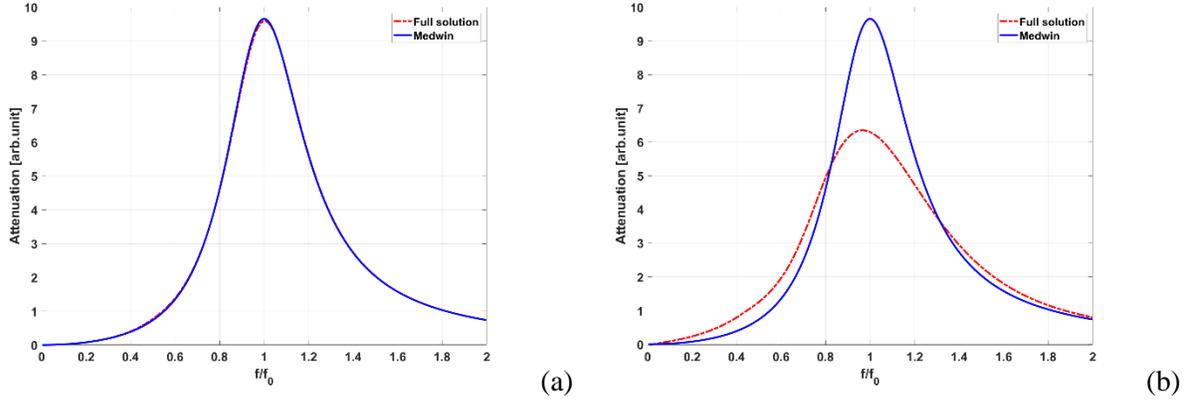

**Figure 2**: Comparison of the attenuation curves estimated by the full solution (red dash-dot) and Medwin's (blue solid) formula vs. excitation frequency. (a) from steady state oscillation, (b) from transient oscillation.

The above result is not new to the theory of harmonic oscillator. However, the key point is the impact on the material characterization of UCAs. The shell parameters of an UCA is often determined by attenuation data that are measured by broadband ultrasonic transducers at low excitation pressures (Frinking and de Jong 1998, Hoff, Sontum et al. 2000, Gong, Cabodi et al. 2014, Raymond, Haworth et al. 2014). Most of the broadband transducers used in the attenuation measurement have a pulse duration of less than 1 µs. For instance, the pulse duration of a 2.25 MHz unfocused transducer is about 1 / 2.25 µs. This indicates that the UCA (see **Figure 1**) has not yet rung up to steady state oscillation before the driven pressure terminates. Therefore, material characterization based on the Medwin's attenuation formula could result in inaccurate estimation of shell parameters. When these shell parameters are employed to predict the nonlinear dynamics of the UCA, such as subharmonics and ultraharmonics generations, misinterpretation could also occur.

### III. ATTENUATION DUE TO TRANSIENT OSCILLATION

Based on the aforementioned discussion, an attenuation formulation considering transient oscillation will be helpful in the material characterization of UCA. Since Eq.(12) is not computational friendly for inversely solving the shell parameters, we obtain an approximated attenuation formula by substituting $R(t) = R_0 + X(t)$ into Eq.(12) and neglecting all the second order terms, which is in the form of

$$\sigma_e = \sigma_e^{st} + \frac{4\omega\rho c_0 R_0}{N P_A} \int_0^{2N\pi/\omega} \dot{X}^{tr} \cos(\omega t) dt \qquad (15)$$

where *N* is the number of the period such that the amplitude of the transient solution decreases to 1% of the initial value, which is obtained via Eq.(7) as $N = 2\ln(10)/\pi\delta$. Note that the period should be the pulse duration in a pulse-echo system.

**Figure** *3* displays the attenuation curves in the transient region calculated by the full solution (red dash-dot), Medwin's formula (blue solid), and Eq.(15) (cyan short dash). The difference between the full solution and the approximated formula is less than 8%, suggesting that the latter can be used to represent the attenuation calculated by the full solution.

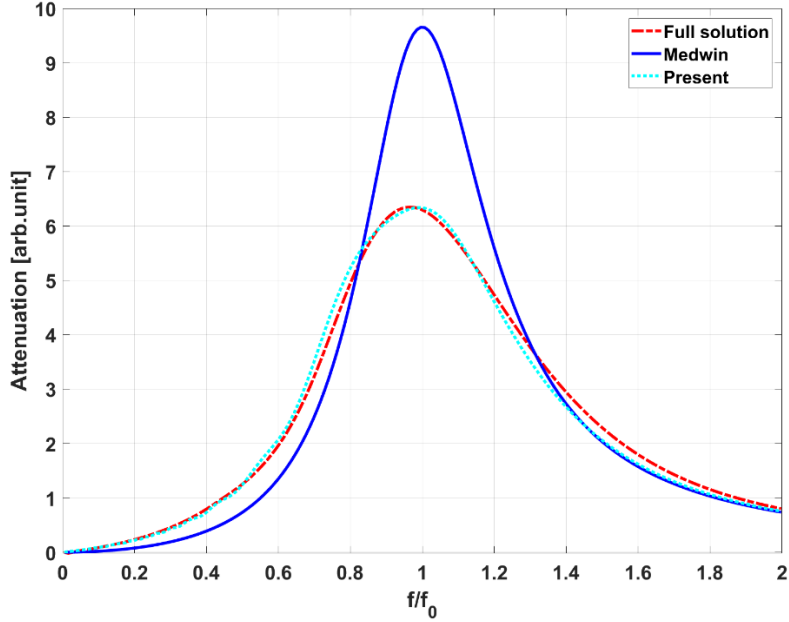

**Figure 3**: The attenuation curves estimated by the full solution (red dash-dot), Medwin's formula (blue solid), and the present calculation (Eq.(15), cyan short-dash).

We can use Eq.(15) as a model equation to estimate the shell parameters of a UCA by fitting the model curve to the experimental attenuation data (Xia, Porter et al. 2015). In this modeling, the excitation pressure was 40 kPa, and the reference surface tension was kept as 0.04 N/m. **Figure** *4* shows the model curves that best match the experimental data (circles), of which the estimated dilatational viscosities and elasticities are listed in **Table** *1*. While the dilatation elasticity estimated by Eq.(15) is slightly higher than that of the Medwin's estimation due to the fact that the damped resonance frequency is lower than the resonance frequency of the UCA, the difference in dilatational viscosity is significant. Medwin's formula gives the dilatational viscosity almost three times higher than that of Eq.(15). It is not difficult to figure out that these differences in the shell parameter would also result in totally different dynamical behaviors of the UCA.

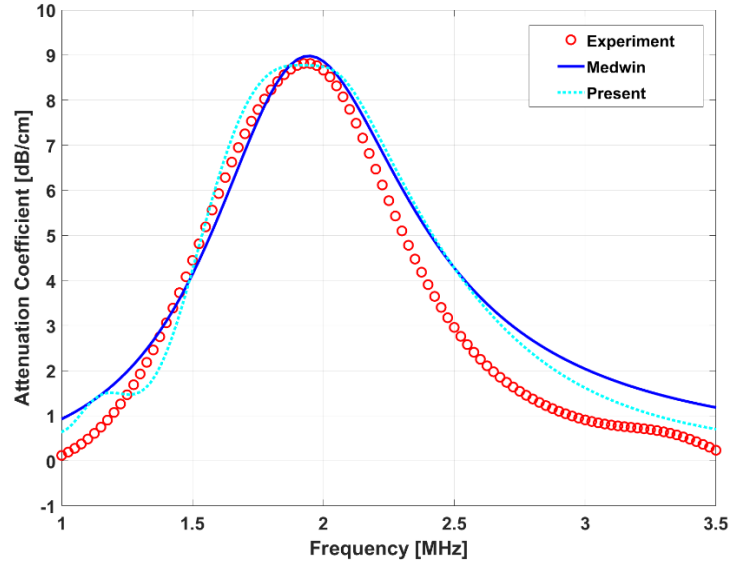

**Figure 4**: The experimental data (Xia, Porter et al. 2015) matched by attenuation curves predicted by Medwin's formula (blue solid), and the present calculation (Eq.(15), cyan short-dash).

**Table 1:** Shell parameters estimated using two different attenuation formulae.

|  | Dilatational viscosity | Dilatational elasticity |
|---|---|---|
| Medwin's(Xia, Porter et al. 2015) | $1.97 \times 10^{-8}$ N.s/m | 0.40 N/m |
| Transient oscillation (Eq.(15)) | $0.62 \times 10^{-8}$ N.s/m | 0.45 N/m |

**IV. CONCLUSION**

Analytical solutions of a linearized Rayleigh-Plesset equation suggest that transient and steady state oscillations of an UCA could be significantly different, which raises concern on the attenuation coefficient calculated by a steady state attenuation formula. Inappropriate estimation of attenuation coefficient could bring uncertainty into the material characterization of the UCA. We thus derived an attenuation formula considering the effect of transient oscillation. It demonstrates that the attenuation coefficient estimated by the new formula near the resonance of the UCA is significantly less than that of estimated by Medwin's formula. The attenuation peak also downshifts to the damped resonance frequency of the UCA. These changes result in alternations of the estimated dilatational viscosities and elasticities, particularly, the dilatational viscosity estimated by the new formula is significantly small. The present results can be a reference when investigating experimental attenuation data measured by short-pulse ultrasound.